\documentclass[11pt, preprint]{article}
\usepackage{color} \usepackage{cancel}
\usepackage{float} \usepackage{graphicx}
\usepackage{subfigure} \usepackage{caption}	
\usepackage{dcolumn}
\usepackage[toc,page]{appendix} \usepackage{authblk}	 		
\usepackage{indentfirst}	
\usepackage{authblk}	 		
\usepackage{multirow}     
\usepackage{hyperref}     
\usepackage{amsfonts, amssymb, amsmath}
\usepackage[left=2.5cm,right=2.5cm,top=2.9cm,bottom=2.9cm,a4paper]{geometry}
\usepackage[noadjust]{cite}
\usepackage{filecontents}
\usepackage{soul}
\usepackage{xcolor}
\setstcolor{red}

\graphicspath{{./figures/}}
\hypersetup{ colorlinks   = true, 
urlcolor     = blue, 
linkcolor    = blue, 
citecolor    = red 	 
}

\title{\Large\bf New method of exponential potentials reconstruction based on given scale factor in phantonical two-field models}
\author[1,2]{I. V. Fomin\thanks{ingvor@inbox.ru}}
\author[1,2,3]{S. V. Chervon\thanks{chervon.sergey@gmail.com}}
\affil[1]{\small \it  Bauman Moscow State Technical University, Russia}
\affil[2]{\small \it Ulyanovsk State Pedagogical University, Ulyanovsk, Russia}
\affil[3]{\small \it Kazan Federal University, Kazan, Russia}

 \date{}
\begin{document}

\maketitle
\begin{abstract}

We investigate two-field cosmological model with phantom and canonical fields (phantonical model as a generalisation of the quintom model for global universe evolution, including early inflationary stage). The model is represented as the chiral cosmological model with the target space conformal to 2D pseudo-Euclidean space. We found three sorts of exact solutions for a constant potential by direct integration
 of dynamic equations and proposed new method of exact solution construction also extended for e-folds N-formalism for the case of non-constant exponential potential. We show that the exact solutions of cosmological dynamic equations can be obtained in explicit form for any type of scale factor evolution $a(t)$ which implies the explicit inverse dependence $t=t(a)$, considering the quasi de Sitter expansion of the universe with non-negligible kinetic energies of scalar fields and showing that the appeared effective cosmological constant can be considered as the source of second accelerated expansion of the universe. Further we analyze cosmological perturbations in the two-field model under consideration reducing it to the single field one. Such transition give us the way of cosmological parameters calculation and comparison them to observational data. We find that in proposed two-field cosmological model the isocurvature perturbations are negligible, and observable curvature perturbations are induced by adiabatic modes only. The series of phantonical models based on exact inflationary solutions are represented, and it is shown the correspondence to observational data for these models.

\end{abstract}

\section{Introduction}

Scalar field cosmology is actively exploited in the  inflation since the very beginning of its
origination \cite{Baumann:2014nda}. Single scalar field with various types of potentials are considering for study early and later inflation \cite{Baumann:2014nda,Chervon:2019sey}.
Multiple scalar fields models are studied as well. It is of interest to mention that additional properties of the models appear when two-field model is under consideration. Such models may lead to various cosmological
properties that were absent in the case of a single scalar field.

Two-field models are studied in different aspects.

False-vacuum decay of a scalar field coupled to a time-dependent external field in the process of bubble nucleation and expansion studied in the work \cite{Widrow91}.
The problem of the cosmological constraints on axion mass in inflationary cosmology, two-field models with inflaton and Peccei-Quinn field, as well as a hybrid inflation with two interacting scalar fields are proposed in the article \cite{Linde91}.
More detailed investigation of the hybrid inflation is presented in the works \cite{Linde94},
\cite{Copel94},
\cite{Kobayashi2015},
\cite{Choi2014}.

Along with the study of the multifield model, special attention is paid to the two-field model with instanton and dilaton field in assisted inflation scenario \cite{Malik99},
\cite{Copel99}.
The splitting into adiabatic and isocurvative perturbations for a model with two scalar fields with kinetic interaction is considered in the work \cite{DiMarco2003}.
The method of soliton solutions construction for the two-field model in 2D Euclidean space is proposed in \cite{Bazeia2003}.
Scalar fields properties for flat galactic rotation curves were considered in the work \cite{Fay2004}
where the two-field model has been studied in static spherically symmetric 
spacetime.
Scaling solutions are investigated within the two-field model in \cite{Tolley07}.
Exact solutions for Euclidean and curved metric of the field space were obtained there.
Other examples of two-field models are presented in \cite{Chervon:2013btx} and literature quoted therein.
%
A complex scalar field with helicoid
potential considered in the work \cite{Li2015}.

Recently the study of chiral cosmological models with two scalar fields and the cosmological constant performed in the work \cite{Vernov2021}.
The two scalar fields model in the framework of phantom inflation investigated in \cite{Nojiri2006,Elizalde:2008yf}.
Two-field scalar-tensor gravity was reviewed in \cite{Elizalde:2004mq,Nojiri2015}.

Two scalar field cosmology where the scalar field interact both in the kinetic part and the potential studied in \cite{Pali2014,Paliathanasis:2018vru,Anguelova:2018vyr}.
Noether point symmetry was investigated for the model and exact solutions were found there.
The superpotential method to construct the exact solutions for multifield cosmological model with interacting scalar fields was considered in \cite{Chervon:2019nwq}.

Let us note about the terminology of a two-field model.
Hybrid inflation \cite{Linde94} with two ordinary (canonical) fields, and quintom model (see \cite{Cai2010} and literature quoted therein) with one quitessence (canonical) field and phantomical one are popular in such models. Quintom model is addressed to a later inflation. When we consider a global universe evolution, including early inflation, the term {\it phantonical} model \cite{Chervon2015}
is more suitable\footnote{The term "phantonical" model \cite{Chervon2015} is constructed by analogy with term "quintom" filed: {\it phanto}mical and cano{\it nical} field.}. So phantonical model covers wider period of time.

Since the universe is expanding at an accelerated rate in present era \cite{SupernovaCosmologyProject:1998vns,SupernovaSearchTeam:1998fmf}, the phenomenologically correct cosmological models should include a description of both stages of the accelerated expansion of the universe.
Various cosmological models are used to describe the first and repeated stages of the accelerated expansion of the universe including such as the cosmological constant, different types of scalar fields and different modifications of General Relativity (see, for example, in \cite{Copeland:2006wr,Clifton:2011jh,Nojiri:2017ncd,Ishak:2018his,
Fomin:2017sbt,Fomin:2019yls,Fomin:2020woj,Fomin:2020hfh}).

In this work we propose a new approach to exact solutions construction for cosmological models based on the doublet of the canonical and phantom scalar fields with negligible kinetic interaction between them. Obtained  solutions capable to describe both stages of the universe's accelerated expansion.
The main feature of the cosmological models for such class of exact solutions is that they admit approximated quasi de Sitter regime with the effective cosmological constant at the large times. %
Thus, the first inflationary stage can be described by two scalar fields with the potential $V\neq const$. However, in contrast to quintom models, the second stage of accelerated expansion of the universe inspired by effective cosmological constant $\Lambda_{eff}$ corresponding to these two fields at times significantly exceeding the time of the first inflationary stage.

The paper is organized as follows. In Sec. \ref{intro} we obtain the cosmological dynamic equations for the two-field phantonical model. In Sec. \ref{V0} the class of exact solutions corresponding to the constant potential $V=const.$, which are reduced to one-filed models, is considered. In Sec. \ref{VEXACT} new method of exact solutions construction for inflationary models with non-constant potential $V\neq const$ is represented.
In Sec. \ref{E-FOLDS} the exact solutions are represented in terms of e-folds number, that allow us to define
the constant of integration in expression for the potential in explicit form. In Sec. \ref{restrictions} the restrictions on the models parameters from initial values  of cosmological dynamics are formulated. In Sec. \ref{cosmconst} the model of effective cosmological constant inspired by the doublet of the canonical and phantom scalar fields in quasi de Sitter regime of accelerated expansion of the universe is studied.
 Sec. \ref{perturbations} is devoted to the analysis of cosmological perturbations.  It is proved that in proposed two-field cosmological models the isocurvature perturbations are negligible, and observable curvature perturbations are induced by adiabatic modes only.
 In Sec. \ref{EXACT-CC} we study various types of universe evolution, including power-law, generalized hyperbolic and exponential power-low inflation. The exact solutions for such inflationary models are obtained and the correspondence to the $\Lambda$CDM--model with effective cosmological constant at the large times is considered. Inflationary parameters are obtained, diagrammers for the index of scalar perturbations $n_s$ vs. tensor-to-scalar ration $r$ are displayed, and correspondence to observational data is proved.
 Sec. \ref{conclusion} contains our summary.

\section{Cosmological dynamic equations for phantonical two-field model}\label{intro}

We will consider the phantonical two-field model as the partial case of more general multifield chiral cosmological models, which are used to describe the universe's dynamics at the both stages of accelerated expansion \cite{Chervon:2013btx,Abbyazov:2014hea} and to construct the models of the black holes and wormholes as well \cite{Chervon:2020kdv,Bronnikov:2009hn}.

Chiral cosmological models with $K$ scalar fields $\phi^A ~ (\bar{\phi}=\phi^1,\phi^2..., \phi^K) $ in the system of units $c=8\pi G=M^{-2}_{p}=1$ are described by the following action \cite{Chervon:2019nwq}
\begin{equation}
S=\int d^4x \sqrt{-g}\left[\frac{1}{2}R-
\frac{1}{2}h_{AB}(\bar{\phi})\partial_\mu\phi^A\partial_\nu\phi^Bg^{\mu\nu}
	-V(\bar{\phi})\right], \label{action}
\end{equation}
where $V(\bar{\phi})$ is the interaction potential and $h_{AB}(\bar{\phi})$ is the metric of a target (a chiral) space.

Dynamic equations for the action (\ref{action}) in a spatially flat Friedmann-Robertson-Walker spacetime can be represented as follows
\begin{equation}
\label{CCM1}
V(\bar{\phi})=3H^{2}+\dot{H},
\end{equation}
\begin{equation}
\label{CCM2}
\frac{1}{2}h_{AB}(\bar{\phi}){\dot\phi}^A{\dot\phi}^B=-\dot{H},
\end{equation}
\begin{equation}
\label{CCM3}
-h_{CB}\left(\ddot{\phi}^B+3H\dot{\phi}^B\right) - h_{CB,D}\dot{\phi}^D\dot{\phi}^B
+ \frac{1}{2}h_{DB,C}\dot{\phi}^D\dot{\phi}^B - V_{,C} = 0,
\end{equation}
where the Hubble parameter $H=\dot{a}/a$. Over dot means the derivative with respect to cosmic time $t$.

In the case of cosmological models without kinetic and cross interaction between scalar fields $\bar{\phi}$ we have to choose
the diagonal metric of a target space $h_{AB}$ with constant components.
Thus, we consider phantonical\footnote{We use the term "pantonical"\ (phantom-canonical) model \cite{Chervon2015} instead of "quintom"\ (quintessence-phantom) model. Quintom model is addressed to a later inflation while "phantonical"\ one can be applied for global universe evolution including the early inflation.} inflationary models with canonical $\phi^{1}=\phi$ and phantom $\phi^{2}=\chi$
fields, with negligible kinetic and cross interaction between them.
For such a model we have the following metric of a target space
\begin{equation}\label{ts-met}
h_{AB}=\omega
\begin{pmatrix}
1 & 0\\
0 & -1\\
\end{pmatrix}=
\begin{pmatrix}
\omega & 0\\
0 & -\omega\\
\end{pmatrix}
,~~(A,B,...=1,2),
\end{equation}
where $\omega>0$ is the conformal factor which defines the dilations of a target space.

For the target metric (\ref{ts-met}) the dynamic equations (\ref{CCM1})--(\ref{CCM3}) are reduced to the following ones
\begin{equation}
\label{CCM1A}
V(\phi,\chi)=3H^{2}+\dot{H}=2\left(\frac{\dot{a}}{a}\right)^{2}+\frac{\ddot{a}}{a},
\end{equation}
\begin{equation}
\label{CCM2A}
\frac{\omega}{2}\dot{\phi}^{2}-\frac{\omega}{2}\dot{\chi}^{2}=-\dot{H},
\end{equation}
\begin{equation}
\label{CCM3A}
\ddot{\phi}+3H\dot{\phi}+\frac{1}{\omega}\frac{\partial V}{\partial\phi}=0,
\end{equation}
\begin{equation}
\label{CCM31A}
\ddot{\chi}+3H\dot{\chi}-\frac{1}{\omega}\frac{\partial V}{\partial\chi}=0.
\end{equation}

We are considering two classes of exact solutions, setting $V=const$ or $V\neq const$,  in the dynamic equations (\ref{CCM1A})--(\ref{CCM31A}).

Let us note that the exact solutions for the dynamic equations (\ref{CCM1A})--(\ref{CCM31A}) with $\omega=1$ were considered earlier in \cite{Vernov:2006dm}, where the superpotential method was applied. Within the framework of such approach, the evolution of scalar fields $\phi=\phi(t)$, $\chi=\chi(t)$ is specified and the cosmological evolution of a scale factor $a=a(t)$ is determined. Further the potential of interaction $V=V(\phi,\chi)$ as the function of the fields (corresponding to given evolution of scalar fields) is reconstructed.
On the contrary, in the present work we set the type of cosmological evolution of a scale factor $a=a(t)$. It allows us to reconstruct the evolution of scalar fields $\phi=\phi(t)$, $\chi=\chi(t)$ and the potential $V=V(\phi,\chi)$  from equations (\ref{CCM1A})--(\ref{CCM31A}). In addition, our method of construction exact solutions leads to the class of two-field cosmological models describing both stages of the accelerated expansion of the universe.

\section{Exact solutions for a constant potential: $V=const.$}\label{V0}

First of all, let us find the exact solutions for the constant potential of scalar fields:
$V=V_*=const.$ From (\ref{CCM3A})--(\ref{CCM31A}) it follows the 
relation between scalar fields:
\begin{equation}
\label{condition}
\phi(t)=n\chi(t)+c,
\end{equation}
where $n$ and $c$ are constants. For the sake of simplicity we define
\begin{equation}
\label{condition1}
\omega-\frac{\omega}{n^{2}}=k.
\end{equation}

Let us note that translation of a scalar field $ \phi \rightarrow \phi + const.$ does not change the model's equations.

Taking into account (\ref{condition}) and (\ref{condition1}),
the equations (\ref{CCM1A})--(\ref{CCM31A}) are reduced to the dynamic system for a single field:

\begin{equation}
\label{CCM1A-1}
V_*=3H^{2}+\dot{H},
\end{equation}
\begin{equation}
\label{CCM2A-1}
\frac{1}{2}\dot{\phi}^{2}k=-\dot{H},
\end{equation}
\begin{equation}
\label{CCM3A-1}
\ddot{\phi}+3H\dot{\phi}=0.
\end{equation}

Further we find the Hubble parameter $H(t)$ from  (\ref{CCM1A-1}), then we define the scalar field $\phi$ solving (\ref{CCM3A-1}), and finally we check  (\ref{CCM2A-1}) to get restriction on integration constants.
The result is as follows.

1. Exponential rate of a scale factor

\begin{equation}
\label{EXP}
H=\lambda =const.,~~~a(t)=a_0 e^{\lambda t},~~V_*=3\lambda^{2},~~k \neq 0,~~\phi=const.,
\end{equation}

\begin{equation}
\label{EXP-2}
H=\lambda,~~~a(t)=a_0 e^{\lambda t},~~V_*=3\lambda^{2},~~k=0,~~\phi=-\frac{\tilde{c}}{3\lambda}e^{-3\lambda t}+\phi_*,
\end{equation}
where $a_{0}$ is the scale factor at the beginning of inflation.

2. Expansion defined by hyperbolic functions
\begin{eqnarray}
\label{COSH}
&&H(t)=\lambda\tanh(3\lambda t),~~~~~a(t)=a_0\cosh^{1/3}(3\lambda t),\\
&&V_*=3\lambda^{2},~~\phi(t)= \sqrt{-\frac{2}{3k}}\arcsin\left(\tanh\left(3\lambda t\right)\right)=
\sqrt{-\frac{2}{3k}}\arctan\left(\sinh (3\lambda t)\right).
\end{eqnarray}
where $k<0$, and

\begin{eqnarray}
\label{SINH}
&&H(t)=\lambda\coth(3\lambda t),~~~a(t)=a_0\sinh^{1/3}(3\lambda t),\\
&&V_*=3\lambda^{2},~~~\phi(t)= \sqrt{\frac{2}{3k}}\ln\left[\tanh\left(\frac{3\lambda}{2}t\right)\right],
\end{eqnarray}
where $k>0$.

3. Expansion defined by trigonometric functions

\begin{eqnarray}
\label{COS}
&&H(t)=-\lambda\tan(3\lambda t),~~~~~a(t)=a_0\cos^{1/3}(3\lambda t),\\
\label{Vcos}
&&V_*=-3\lambda^{2},~~~\phi(t)=\frac{1}{\sqrt{6k}}\ln\left|\frac{1+ \sin(3\lambda t)}{1-\sin(3\lambda t)}\right|,
\end{eqnarray}
where $k>0$.

The solution (\ref{COS})-(\ref{Vcos}) can be represented in another form:
\begin{eqnarray}
\label{SIN}
&&H(t)=\lambda\cot(3\lambda t),~~~a(t)=a_0\sin^{1/3}(3\lambda t),\\
&&V_*=-3\lambda^{2},~~~\phi(t)=-\sqrt{\frac{2}{3k}}{\rm arctanh}\left(\cos(\lambda t)\right)=-\frac{1}{\sqrt{6k}}\ln\left|\frac{1+\cos(3\lambda t}{1-\cos(3\lambda t)}\right|,
\end{eqnarray}
where $k>0$.

Note, that the second field $\chi$ is defined by (\ref{condition}).

Analyzing the dynamics for the models with $V\neq const.$ we must use another approach because of complicity of solving eq. (\ref{CCM1A}).

\section{New method of exact solutions construction for non-constant potential: $V\neq const$}\label{VEXACT}

To construct the exact cosmological solutions of eqs. (\ref{CCM1A})--(\ref{CCM31A}) for the case $V\neq0$ we fix the form of scalar fields satisfying equation (\ref{CCM2A}) as
\begin{eqnarray}
\label{PH1}
&&\phi(t)=\frac{\alpha}{\sqrt{2\omega}}\ln\left(\frac{A}{a_{0}}\,\frac{a^{2}}{\dot{a}}\right),\\
\label{PS1}
&&\chi(t)=\frac{\beta}{\sqrt{2\omega}}\ln\left(\frac{B}{a_{0}}\,\dot{a}\right),
\end{eqnarray}
where $\alpha=\pm1$, $\beta=\pm1$, $A$ and $B$ are positive non-zero constants.

Using substitution
the eqs. (\ref{PH1})--(\ref{PS1}) into (\ref{CCM3A})--(\ref{CCM31A}) we obtain
\begin{eqnarray}
\label{V1}
&&\frac{1}{\omega}\frac{\partial V}{\partial\phi}=-\ddot{\phi}-3H\dot{\phi}=
\frac{\alpha}{\sqrt{2\omega}}\left[\frac{\dddot{a}}{\dot{a}}-\left(\frac{\ddot{a}}{\dot{a}}\right)^{2}
+\frac{\ddot{a}}{a}-4\left(\frac{\dot{a}}{a}\right)^{2}\right],\\
\label{V2}
&&\frac{1}{\omega}\frac{\partial V}{\partial\chi}=\ddot{\chi}+3H\dot{\chi}=
\frac{\beta}{\sqrt{2\omega}}\left[\frac{\dddot{a}}{\dot{a}}-\left(\frac{\ddot{a}}{\dot{a}}\right)^{2}+
3\frac{\ddot{a}}{a}\right].
\end{eqnarray}
In addition, from eqs. (\ref{CCM1A}) and (\ref{V1})--(\ref{V2}) one can derive the equation\footnote{The  constants $\alpha=1/\alpha$ and $\beta=1/\beta$ by their definition.}
\begin{equation}
\label{V3}
V(\phi,\chi)=\frac{1}{\sqrt{2\omega}}\left(\beta\frac{\partial V(\phi,\chi)}{\partial\chi}-
\alpha\frac{\partial V(\phi,\chi)}{\partial\phi}\right).
\end{equation}
The solution of eq. (\ref{V3}) is
\begin{equation}
\label{V4}
V(\phi,\chi)=f_{1}\left(\frac{\phi}{\alpha}+\frac{\chi}{\beta}\right)
\exp\left(-\frac{\sqrt{2\omega}}{\alpha}\phi\right)+
f_{2}\left(\frac{\phi}{\alpha}+\frac{\chi}{\beta}\right)
\exp\left(\frac{\sqrt{2\omega}}{\beta}\chi\right).
\end{equation}
It is easy to show that the solution (\ref{V4}) can be reduced without loss of generality to the following form
\begin{equation}
\label{V5}
V(\phi,\chi)=f\left(\frac{\phi}{\alpha}+\frac{\chi}{\beta}\right)
\exp\left(-\frac{\sqrt{2\omega}}{\alpha}\phi\right),
\end{equation}
where $f=f\left(\frac{\phi}{\alpha}+\frac{\chi}{\beta}\right)$ is an
arbitrary function of the combination $\left(\frac{\phi}{\alpha}+\frac{\chi}{\beta}\right)$ as the argument.

Using the equations (\ref{CCM1A}) and (\ref{V5}), we can represent the function $f=f\left(\frac{\phi}{\alpha}+\frac{\chi}{\beta}\right)$ as
\begin{eqnarray}
\label{V6}
\nonumber
&&f\left(\frac{\phi}{\alpha}+\frac{\chi}{\beta}\right)=
\left[2\left(\frac{\dot{a}}{a}\right)^{2}+\frac{\ddot{a}}{a}\right]
\exp\left(\frac{\sqrt{2\omega}}{\alpha}\phi\right)\\
&&=\left(\frac{A}{a_{0}}\right)\frac{a^{2}}{\dot{a}}
\left[2\left(\frac{\dot{a}}{a}\right)^{2}+\frac{\ddot{a}}{a}\right]=
\frac{A}{a_{0}}\left[2\dot{a}+\frac{a\ddot{a}}{\dot{a}}\right].
\end{eqnarray}

From expressions (\ref{PH1})--(\ref{PS1}) one can find
\begin{equation}
\label{V6A}
\sqrt{\frac{\omega}{2}}\left(\frac{\phi}{\alpha}+\frac{\chi}{\beta}\right)=\ln a+\frac{1}{2}\ln\left(\frac{AB}{a^{2}_{0}}\right).
\end{equation}
Thus, from eq. (\ref{V6A}) we find the dependence of the scale factor on the fields as
\begin{equation}
\label{V6B}
a(\phi,\chi)=\frac{a_{0}}{\sqrt{AB}}
\exp\left(\sqrt{\frac{\omega}{2}}\frac{\phi}{\alpha}+\sqrt{\frac{\omega}{2}}\frac{\chi}{\beta}\right).
\end{equation}

Further, we consider the following function
\begin{equation}
\label{V7}
f\left(\frac{\phi}{\alpha}+\frac{\chi}{\beta}\right)\equiv
\tilde{f}\left[\exp\left(\sqrt{\frac{\omega}{2}}\frac{\phi}{\alpha}+
\sqrt{\frac{\omega}{2}}\frac{\chi}{\beta}\right)\right]=
const\times \tilde{f}(a).
\end{equation}
The function $\tilde{f}$ is defined by the equation
\begin{equation}
\label{V8}
\tilde{f}(a)=2\dot{a}+\frac{a\ddot{a}}{\dot{a}}.
\end{equation}

With the definition (\ref{V8}) the potential (\ref{V5}) becomes
\begin{equation}
\label{V9}
V(\phi,\chi)=const\times\tilde{f}
\left[\exp\left(\sqrt{\frac{\omega}{2}}\frac{\phi}{\alpha}
+\sqrt{\frac{\omega}{2}}\frac{\chi}{\beta}\right)\right]
\exp\left(-\frac{\sqrt{2\omega}}{\alpha}\phi\right).
\end{equation}
I.e. we generalized exponential potential corresponding to scalar fields (\ref{PH1})--(\ref{PS1}).

Thus, one can find the exact cosmological solutions of equations (\ref{CCM1A})--(\ref{CCM31A}) for any scale factor implying the function $\tilde{f}=\tilde{f}(a)$ in explicit form. Such function can be obtained from equation (\ref{V8}) based on the inverse dependence $t=t(a)$ for the scale factor $a=a(t)$ under consideration.

Now, we will consider the exact solutions in terms of e-folds number to find the constant in the expression (\ref{V9}) for the potential in explicit form.

\section{The method of exact solutions construction in terms of e-folds number $N$}\label{E-FOLDS}

As additional characteristic of inflationary dynamics we consider the e-folds number $N$, which is defined as follows
\begin{eqnarray}
\label{EX1}
&&H=\dot{N},\\
\label{EX2}
&&N=\ln\left(\frac{a}{a_{0}}\right),\\
\label{EX3}
&&a(t)=a_{0}\exp\left(N(t)\right).
\end{eqnarray}

Now, we rewrite equation (\ref{V8}) as
\begin{equation}
\label{EX4}
\tilde{f}(a)=a\left(3H+\frac{\dot{H}}{H}\right)=a\left(3\dot{N}+\frac{\ddot{N}}{\dot{N}}\right).
\end{equation}
In terms of the e-folds number $N$, equation (\ref{EX4}) can be rewritten as
\begin{eqnarray}
\label{EX5}
&&\tilde{f}(a)\propto\exp(N)F(N),\\
\label{EX6}
&&F(N)=3\dot{N}+\frac{\ddot{N}}{\dot{N}},
\end{eqnarray}
where $F(N)$ is a function of e-folds number $N$.

In a similar way, the scalar fields (\ref{PH1})--(\ref{PS1}) can be rewritten in terms of e-folds number as
\begin{eqnarray}
\label{EX7}
&&\phi(t)=\frac{\alpha}{\sqrt{2\omega}}\left[N(t)-\ln\dot{N}+\ln A\right],\\
\label{EX8}
&&\chi(t)=\frac{\beta}{\sqrt{2\omega}}\left[N(t)+\ln\dot{N}+\ln B\right].
\end{eqnarray}

Combining equations (\ref{EX7})--(\ref{EX8}) one can obtain the following
\begin{eqnarray}
\label{EX9}
&&N(\phi,\chi)=\sqrt{\frac{\omega}{2}}\frac{\phi}{\alpha}
+\sqrt{\frac{\omega}{2}}\frac{\chi}{\beta}-\frac{1}{2}\ln(AB),\\
\label{EX10}
&&-\sqrt{\frac{\omega}{2}}\frac{\phi}{\alpha}
+\sqrt{\frac{\omega}{2}}\frac{\chi}{\beta}=\ln \dot{N}+\frac{1}{2}\ln\left(\frac{B}{A}\right).
\end{eqnarray}

Using (\ref{V7}), (\ref{V9}), and (\ref{EX5}),  the potential can be represented as
\begin{eqnarray}
\label{EX11}
V(\phi,\chi)=V_{0}\exp\left(\sqrt{\frac{\omega}{2}}\frac{\phi}{\alpha}
+\sqrt{\frac{\omega}{2}}\frac{\chi}{\beta}\right)
\exp\left(-\frac{\sqrt{2\omega}}{\alpha}\phi\right)\times F(N),
\end{eqnarray}
where $V_{0}$ is a constant, i.e. the solution for the potential is
\begin{equation}
\label{EX12}
V(\phi,\chi)=V_{0}\exp\left(-\sqrt{\frac{\omega}{2}}\frac{\phi}{\alpha}
+\sqrt{\frac{\omega}{2}}\frac{\chi}{\beta}\right)\times F\left(N\right).
\end{equation}

At the other hand, the potential (\ref{CCM1A}) can be displayed as
\begin{equation}
\label{EX13}
V(t)=3H^{2}+\dot{H}=3\dot{N}^{2}+\ddot{N}=F(N)\dot{N},
\end{equation}
and, taking into account (\ref{EX10}), from eqs. (\ref{EX12}) and (\ref{EX13}), we obtain
\begin{equation}
\label{EX14}
V_{0}=\sqrt{\frac{A}{B}}\,.
\end{equation}

Thus, we represent the potential in the final form as
\begin{equation}
\label{EX15}
V(\phi,\chi)=\sqrt{\frac{A}{B}}\exp\left(-\sqrt{\frac{\omega}{2}}\frac{\phi}{\alpha}
+\sqrt{\frac{\omega}{2}}\frac{\chi}{\beta}\right)\times F\left(N(\phi,\chi)\right),
\end{equation}
where $N(\phi,\chi)$ is defined by eq. (\ref{EX9}).

Now, we consider restrictions on the model parameters from initial values of cosmological dynamics.

\section{The restrictions on the model parameters of cosmological dynamics}\label{restrictions}

In view of standard cosmology we request that at the time $t_{i}$ of the beginning of inflation the relations
\begin{equation}
\label{CR}
a(t=t_{i})=a_{0},~~~N(t=t_{i})=0
\end{equation}
should be satisfied.

To have finite values of the fields at the beginning of inflation
we have to impose from eqs. (\ref{EX7})--(\ref{EX8}) the following condition
\begin{equation}
\label{CR1}
\dot{N}(t=t_{i})\neq0.
\end{equation}
Evidently, when this condition is violated, one obtains divergences $|\phi| \rightarrow\infty$ and $|\chi| \rightarrow\infty$ at the time $t=t_{i}$.

In a similar way, using eq. (\ref{EX15}), we determine the second condition
\begin{equation}
\label{CR2}
F(N=0)\neq\pm\infty .
\end{equation}
When this condition is violated, we have divergence $V\rightarrow\pm\infty$ at the beginning of inflation.

Thus, we can formulate the following assertion for the special class of the exact cosmological solutions under consideration:

{\it The exact solutions of cosmological dynamic equations (\ref{CCM1A})--(\ref{CCM31A}) can be obtained in explicit form for any type of scale factor evolution $a=a(t)$ (evolution of e-folds number $N=N(t)$) which implies the explicit inverse dependence $t=t(a)$ (inverse dependence $t=t(N)$) and satisfy conditions (\ref{CR1})--(\ref{CR2}).}

Also, we note, that condition (\ref{CR1}) is satisfied for expansion of the early universe $\dot{N}(t=t_{i})=H(t=t_{i})>0$.
Since, during inflation, the e-folds number growing from $N=0$ to $N\simeq 60$, one has $\ddot{N}>-\dot{N}^2$ for any inflationary model implying accelerated expansion of the early universe.
Thus, the condition $\ddot{N}>0$ at the inflationary stage
is sufficient to get relevant cosmological models implying accelerated expansion of the universe.

Now, we consider the application of proposed approach to the analysis of the phantonical two-field cosmological models.

\section{The effective cosmological constant}\label{cosmconst}

According to the inflationary paradigm, the early universe is subject to accelerated expanding in the regime close to exponential one $a(t)\simeq a_{0}\exp(\lambda t)$. Such expansion corresponds to the quasi de Sitter stage, which is equivalent to conditions $H\simeq\lambda$ and $\dot{H}\simeq0$. Became known, according to the modern observations, that the universe is currently going through an accelerated phase of expansion as well. Besides, the acceleration of the universe expanding at the first inflationary stage is significantly larger than the acceleration during the second (later) inflation $\ddot{a}_{inf}\gg \ddot{a}_{sec}$ \cite{SupernovaCosmologyProject:1998vns,SupernovaSearchTeam:1998fmf}.

Let us analyze the solutions in Sec. \ref{E-FOLDS}  for the case of an accelerated expansion of the universe closely to the de Sitter stage based on the condition $\dot{H}\simeq0$.

For the models with one canonical scalar field ($\chi=0$) using eq. (\ref{CCM2A}), the condition $\dot{H}\simeq0$ leads to $X^{(\phi)}\equiv\frac{\omega}{2}\dot{\phi}^{2}\simeq 0$. Therefore, conditions $V\gg X^{(\phi)}$ and $\ddot{\phi}\simeq0$ correspond to negligible kinetic energy and to negligible acceleration of a scalar field $\phi$ during inflationary  stage in slow-roll regime \cite{Baumann:2014nda,Chervon:2019sey}.

As for the model under consideration in Sec. \ref{VEXACT}, the kinetic energies of scalar fields $\phi$ and $\chi$ are not restricted by the slow-roll regime, since their substitution from (\ref{PH1})--(\ref{PS1}) into (\ref{CCM2A}) imply the satisfaction of any condition on $\dot{H}$ including $\dot{H}\simeq0$.
Therefore, we will consider approximate solutions for phantonical two-fields inflationary models under condition $\dot{H}\simeq0$ only, corresponding to the quasi de Sitter expansion of the universe with non-negligible (in general case) kinetic energies of the scalar fields.

Based on the condition $H\simeq\lambda$ we have
\begin{equation}
\label{A}
N(t)\simeq \lambda t.
\end{equation}
Using the dependence (\ref{A}), from eqs. (\ref{EX7})--(\ref{EX8}) we obtain the linear evolution of the scalar fields:
\begin{eqnarray}
\label{A1}
&&\phi(t)\simeq\frac{\alpha}{\sqrt{2\omega}}\left[\lambda t+\ln\left(\frac{A}{\lambda}\right)\right],\\
\label{A2}
&&\chi(t)\simeq\frac{\beta}{\sqrt{2\omega}}\left[\lambda t+\ln\left(\lambda B\right)\right].
\end{eqnarray}

Thus, in the quasi de Sitter expansion of the universe, using (\ref{A1})--(\ref{A2}), we can obtain the following kinetic energies of the scalar fields
\begin{equation}
\label{KIN}
X^{(\phi)}=-X^{(\chi)}=\frac{\omega}{2}\dot{\phi}^{2} \simeq\frac{\lambda^{2}}{4},~~X^{(\phi)}= \frac{1}{2}h_{11}\dot{\phi}^{2}, ~~X^{(\chi)}= \frac{1}{2}h_{22}\dot{\chi}^{2},
\end{equation}
and the total kinetic energy
\begin{equation}
\label{KIN2}
X^{(\phi,\chi)}=\frac{1}{2}h_{11}\dot{\phi}^{2}+ \frac{1}{2}h_{22}\dot{\chi}^{2} =X^{(\phi)}-X^{(\chi)}\simeq0.
\end{equation}

To get an approximated expression of the potential (\ref{EX15}) we find $F(N)\simeq3\lambda$ from (\ref{EX6}) using (\ref{A}). As the result we have
\begin{equation}
\label{A3}
V(\phi,\chi)\simeq3\lambda\sqrt{\frac{A}{B}}\exp\left(-\sqrt{\frac{\omega}{2}}\frac{\phi}{\alpha}
+\sqrt{\frac{\omega}{2}}\frac{\chi}{\beta}\right).
\end{equation}

Substituting approximated expressions for scalar fields evolution (\ref{A1})--(\ref{A2}) into (\ref{A3}), we obtain
\begin{equation}
\label{A4}
V\simeq3\lambda^{2}.
\end{equation}
Thus, for the quasi de Sitter expansion of the universe we have
\begin{equation}
\label{A5}
\frac{V}{X^{(\phi)}}=\frac{V}{|X^{(\chi)}|}\simeq 12.
\end{equation}

It is important to note, that the slow-roll condition $V\gg X^{(\phi,\chi)}$ is fulfilled for non-negligible constant kinetic energies of the scalar fields, and corresponding state parameter is
\begin{equation}
\label{A6}
w=\frac{X^{(\phi,\chi)}-V}{X^{(\phi,\chi)}+V}\simeq-1.
\end{equation}

As one can see, solutions (\ref{A1})--(\ref{A2}) and (\ref{A4}) satisfy the conditions (\ref{CR1})--(\ref{CR2}), since $F\simeq 3\dot{N}=3\lambda=const.$ and $\dot{N}\simeq const.$

It should be emphasized here, that solutions (\ref{A1})--(\ref{A2}) and (\ref{A4}) are approximate ones and these solutions do not coincide with exact solutions (\ref{EXP})--(\ref{EXP-2}) obtained for $V=const$.
Therefore, the approximated solutions of the dynamic equations for the doublet of the canonical and phantom scalar fields (\ref{PH1})--(\ref{PS1}) in quasi de Sitter regime correspond to the effective cosmological constant $\Lambda_{eff}=V\simeq3\lambda^{2}$, which is defined by linear dependence of the fields evolution (\ref{A1})--(\ref{A2}).

Thus, the exact solutions of dynamic equations (\ref{CCM1A})--(\ref{CCM31A}) with $V\neq const$ correspond to the deviation from quasi de Sitter model at the first inflationary stage of the universe evolution.
Nevertheless, extending the model for global evolution, the effective cosmological constant $\Lambda_{eff}$ can be considered as the source of the second accelerated expansion of the universe, when the exact inflationary solutions are reduced to (\ref{A1})--(\ref{A2}) and (\ref{A4}) at the large times $t\gg t_{i}$.

We also note that the potential $V(\phi,\chi)$ is defined in dynamic equations (\ref{CCM1A})--(\ref{CCM31A}) up to a constant $V(\phi,\chi)\rightarrow V(\phi,\chi)+\Lambda_{vac}$ associated with the quantum zero point fluctuations~\cite{Padmanabhan:2002ji,Martin:2012bt} corresponding to positive or negative constant $\Lambda_{vac}$.
So, the value of the cosmological constant at present era can be defined as follows
\begin{equation}
\label{CCPE}
\Lambda_{obs}=\Lambda_{eff}+\Lambda_{vac}.
\end{equation}
Therefore, one can estimate the cosmological constant  associated with the vacuum energy $\Lambda_{vac}$ in the models under consideration from expression (\ref{CCPE}) for known values $\Lambda_{obs}$ and $\Lambda_{eff}$.

\section{The cosmological perturbations}\label{perturbations}

 We turn, now, to the analysis of the cosmological perturbations in the inflationary models under consideration.
In multifield cosmological models there are two types of cosmological perturbations, namely adiabatic and isocurvature ones in contrast to the single field models which imply adiabatic perturbations only \cite{Starobinsky:2001xq,Wands:2002bn,Kaiser:2012ak,Kaiser:2013sna,Guerrero:2020lng}.

The usual way to calculate the parameters of the curvature perturbations $\mathcal R$ corresponding to adiabatic modes is reducing the multifield models to the single field ones \cite{Kaiser:2012ak,Kaiser:2013sna,Guerrero:2020lng}.

For the phantonical cosmological models under consideration the action is
\begin{equation}
S_{ph}=\int d^4x \sqrt{-g}\left[\frac{1}{2}R-
\frac{\omega}{2}g^{\mu\nu}\partial_\mu\phi\partial_\nu\phi
+\frac{\omega}{2}g^{\mu\nu}\partial_\mu\chi\partial_\nu\chi
	-V(\phi,\chi)\right]. \label{action1}
\end{equation}

The action (\ref{action1}) can be reduced to the single field action,
\begin{equation}
S_{single}=\int d^4x \sqrt{-g}\left[\frac{1}{2}R-
\frac{\gamma}{2}g^{\mu\nu}\partial_\mu\sigma\partial_\nu\sigma
	-V(\sigma)\right], \label{action2}
\end{equation}
for new scalar field $\sigma$, which is defined by the relation
\cite{Chervon2003gc}:
\begin{equation}
\label{P1}
\frac{1}{2}\dot{\sigma}^{2}=\frac{1}{2}h_{AB}\dot{\phi}^{A}\dot{\phi}^{B}=\frac{\omega}{2}\dot{\phi}^{2}-
\frac{\omega}{2}\dot{\chi}^{2}.
\end{equation}
 Corresponding dynamic equations \cite{Baumann:2014nda,Chervon:2019sey} are
\begin{eqnarray}
\label{P2}
&&V(\sigma)=3H^{2}+\dot{H},\\
\label{P3}
&&\gamma\dot{\sigma}^{2}=-2\dot{H},\\
\label{P4}
&&\ddot{\sigma}+3H\dot{\sigma}+\gamma\frac{dV(\sigma)}{d\sigma}=0,
\end{eqnarray}
where, of the three equations, only two are independent, and $\gamma=\pm1$ corresponds to the canonical ($\gamma=1$ for $\dot{\sigma}^{2}>0$) and phantom ($\gamma=-1$ for $\dot{\sigma}^{2}<0$) fields.

Let us note here that different methods of exact solutions construction of equations (\ref{P2})--(\ref{P4}) are considered in \cite{Chervon:2019sey,Chervon:2017kgn,Fomin:2018uql}.

The parameters of cosmological perturbations in single field inflationary models can be calculated based on the following expressions:

$\bullet$ the power spectra of scalar and tensor perturbations are \cite{Baumann:2014nda,Chervon:2019sey}
\begin{equation}
\label{PSGB}
{\mathcal P}_{S}=A_{S}=\frac{1}{2\gamma\epsilon}\left(\frac{H}{2\pi}\right)^{2},~~~~~~
{\mathcal P}_{T}=2s\left(\frac{H}{2\pi}\right)^{2};
\end{equation}

$\bullet$ the spectral indices of scalar and tensor perturbations, tensor-to-scalar ratio, and the running of the spectral index of scalar perturbations are
\begin{eqnarray}
\label{P5}
&&n_{S}-1=\frac{-4\epsilon+2\delta}{1-\epsilon},\\
\label{P6}
&&n_{T}=-\frac{2\epsilon}{1-\epsilon},\\
\label{P7}
&&r\equiv\frac{{\mathcal P}_{T}}{{\mathcal P}_{S}}=4s\gamma\epsilon,\\
\label{Run}
&&\frac{dn_s}{d\ln k}=\frac{1}{H(1-\epsilon)}\frac{dn_s}{d t},
\end{eqnarray}
where the slow-roll parameters
\begin{eqnarray}
\label{SR1}
&&\epsilon=\frac{\gamma\dot{\sigma}^{2}}{2H^{2}}=-\frac{\dot{H}}{H^{2}},~~~~~~~
\delta=-\frac{\ddot{\sigma}}{H\dot{\sigma}}=-\frac{\ddot{H}}{2H\dot{H}},
\end{eqnarray}
must satisfy the slow-roll conditions
\begin{eqnarray}
\label{SR2}
&&\epsilon\ll1,~~~~~~~\delta\ll1,
\end{eqnarray}
during inflation.

The parameter $s$ depends on the background cosmology and it estimates of the densities of matter
and cosmological constant at the present time \cite{Kinney:2003uw}.
The commonly used conventional value of this parameter is $s=4$ \cite{Baumann:2014nda}. However, the other estimates can be found in the literature, namely $s=3.45$ in \cite{Noh:2001ia}, $s=2.5$ in \cite{Kinney:2003uw,Piao:2004tq} and $s=1$ in \cite{Hwang:2005hb,Odintsov:2018zhw}. Thus, for the sake of generality, we will consider $s=1$, $s=4$ corresponding to the limit estimates of a given parameter \cite{Fomin:2019yls}.

In situation, when slow-roll conditions $\epsilon\ll1$ and $\delta\ll1$ are fulfilled, one can define the slow-roll parameters in terms of the scalar field potential \cite{Baumann:2014nda,Chervon:2019sey,Piao:2004tq}
\begin{eqnarray}
\label{SRV1}
&&\epsilon\simeq\frac{\gamma}{2}\left(\frac{V'_{\sigma}}{V}\right)^{2}\equiv\epsilon_{V},\\
&&\delta\simeq\frac{V''_{\sigma}}{V}-\frac{\gamma}{2}\left(\frac{V'_{\sigma}}{V}\right)^{2}
\equiv\eta_{V}-\epsilon_{V}.
\end{eqnarray}

Nevertheless, expressions (\ref{PSGB})--(\ref{SR1}) allow one to calculate the parameters of cosmological perturbations based on the Hubble parameter as the function of cosmic time $H=H(t)$ only.

The constraints on the values of the parameters of cosmological perturbations follows from the combination of cosmic microwave background radiation (CMB) and baryon acoustic oscillations (BAO) data~\cite{Planck:2018vyg,BICEP2:2018kqh}
\begin{eqnarray}
\label{PLANCK1}
&&A_{S}=2.1\times10^{-9},~~~~~~~~~~~~~~~n_{S}=0.9663\pm 0.0041,\\
\label{PLANCK2}
&&r<0.1~~~\text{(Planck 2018)},~~~~~~r<0.065~~~\text{(Planck 2018/BICEP2/Keck-Array)},
\end{eqnarray}
restrict the possible models of cosmological inflation.

Since, the condition (\ref{PSGB}) and range $N=50-60$ can be satisfied by the choice of the constant model's parameters and the time of the crossing of the Hubble radius $t=t_{\ast}$, it suffices to consider the dependence of the tensor-to-scalar ratio $r$ on the spectral index of scalar perturbations $n_{S}$ for verification of the inflationary models by the observational constraints on the parameters of cosmological perturbations.

The curvature perturbations $\mathcal R$ remains constant for purely adiabatic perturbations in the large-scale limit, however, the isocurvature perturbations ${\mathcal S}$ can change the curvature perturbations by non-adiabatic pressure perturbations or energy transfer \cite{Starobinsky:2001xq,Wands:2002bn,Kaiser:2012ak,Kaiser:2013sna,Guerrero:2020lng}.
For these reasons it is necessary to determine the influence of the isocurvature perturbations on the values of the parameters of cosmological perturbations.

In order to determine curvature and isocurvature perturbations we reduce the action (\ref{action1}) for phantonical models to the action corresponding to inflationary models with two non-interacting canonical scalar fields
\begin{equation}
S_{can}=\int d^4x \sqrt{-g}\left[\frac{1}{2}R-
\frac{1}{2}g^{\mu\nu}\partial_\mu\tilde{\phi}\partial_\nu\tilde{\phi}
-\frac{1}{2}g^{\mu\nu}\partial_\mu\tilde{\chi}\partial_\nu\tilde{\chi}
	-V(\tilde{\phi},\tilde{\chi})\right], \label{action3}
\end{equation}
based on the connections
\begin{eqnarray}
\label{P8}
&&\tilde{\phi}=\alpha\sqrt{\omega}\phi,\\
\label{P9}
&&\tilde{\chi}=\beta\sqrt{-\omega}\chi.
\end{eqnarray}

The curvature perturbations, when slow-roll conditions (\ref{SR1}) are satisfied, can be defined as follow \cite{Wands:2002bn}
\begin{equation}
\label{CURV}
{\mathcal R}\simeq H\left[\frac{\dot{\tilde{\phi}}\delta\tilde{\phi}+\dot{\tilde{\chi}}\delta\tilde{\chi}}
{\dot{\tilde{\phi}}^{2}+\dot{\tilde{\chi}}^{2}}\right]=
H\left[\frac{\dot{\phi}\delta\phi-\dot{\chi}\delta\chi}{\dot{\phi}^{2}-\dot{\chi}^{2}}\right],
\end{equation}
and isocurvature perturbations are \cite{Wands:2002bn}
\begin{equation}
\label{ISOCURV}
{\mathcal S}= H\left[\frac{\dot{\tilde{\phi}}\delta\tilde{\chi}-\dot{\tilde{\chi}}\delta\tilde{\phi}}
{\dot{\tilde{\phi}}^{2}+\dot{\tilde{\chi}}^{2}}\right]=
\alpha\beta H\left[\frac{\dot{\chi}\delta\phi-\dot{\phi}\delta\chi}{\dot{\phi}^{2}-\dot{\chi}^{2}}\right],
\end{equation}
where $\delta\phi$ and $\delta\chi$ are the perturbations of the scalar fields.

Since at the inflationary stage slow-roll conditions (\ref{SR1}) are satisfied, and the regime of accelerated expansion of the early universe,  based on eqs. (\ref{A1})--(\ref{A2}),
close to de Sitter one, we can define the following relation between the scalar fields during inflation
\begin{equation}
\label{RELDS}
\chi=\frac{\beta}{\alpha}\phi+\theta+const,~~~~~\theta\ll\phi,\chi .
\end{equation}
Here $\theta=\theta(t)$ is a small arbitrary parameter, which characterizes the deviations from de Sitter stage.
Thus, using (\ref{RELDS}) we can write
\begin{eqnarray}
\label{P10}
&&\dot{\chi}=\frac{\beta}{\alpha}\dot{\phi}+\dot{\theta},\\
\label{P11}
&&\delta\chi=\frac{\beta}{\alpha}\delta\phi+\delta\theta.
\end{eqnarray}

Substituting (\ref{P10})--(\ref{P11}) into eq. (\ref{CURV}) we have
\begin{equation}
\label{CURV1}
{\mathcal R}\simeq H\left[\frac{\dot{\phi}\delta\theta+\dot{\theta}\delta\phi+\frac{\alpha}{\beta}\dot{\theta}\delta\theta}
{2\dot{\phi}\dot{\theta}-\frac{\alpha}{\beta}\dot{\theta}^{2}}\right],
\end{equation}
and neglecting small higher-order terms
\begin{equation}
\label{SMALL}
\dot{\theta}\delta\theta\simeq0,~~~~~\dot{\theta}^{2}\simeq0,
\end{equation}
we get
\begin{equation}
\label{CURV2}
{\mathcal R}\simeq \frac{H}{2}\left(\frac{\delta\phi}{\dot{\phi}}+\frac{\delta\theta}{\dot{\theta}}\right).
\end{equation}

In addition, substituting (\ref{P10})--(\ref{P11}) into eq. (\ref{ISOCURV}), under conditions (\ref{SMALL}), we obtain
\begin{equation}
\label{ISOCURV2}
{\mathcal S}\simeq \frac{H}{2}\left(\frac{\delta\theta}{\dot{\theta}}-\frac{\delta\phi}{\dot{\phi}}\right).
\end{equation}

At the other hand, from eq. (\ref{RELDS}) one can define the inverse relations:
\begin{eqnarray}
\label{P12}
&&\dot{\phi}=\frac{\alpha}{\beta}\dot{\chi}-\frac{\alpha}{\beta}\dot{\theta},\\
\label{P13}
&&\delta\phi=\frac{\alpha}{\beta}\delta\chi-\frac{\alpha}{\beta}\delta\theta.
\end{eqnarray}

Substituting (\ref{P12})--(\ref{P13}) into eq. (\ref{CURV})-(\ref{ISOCURV}), under conditions (\ref{SMALL}), we get
\begin{eqnarray}
\label{CURV3}
&&{\mathcal R}\simeq \frac{H}{2}\left(\frac{\delta\chi}{\dot{\chi}}+\frac{\delta\theta}{\dot{\theta}}\right),\\
\label{ISOCURV3}
&&{\mathcal S}\simeq \frac{H}{2}\left(\frac{\delta\theta}{\dot{\theta}}-\frac{\delta\chi}{\dot{\chi}}\right).
\end{eqnarray}

Thus, from eqs. (\ref{CURV2})--(\ref{ISOCURV2}) and (\ref{CURV3})--(\ref{ISOCURV3}) one has the following relation
\begin{eqnarray}
\label{ADIABATIC}
&&\frac{\delta\phi}{\dot{\phi}}\simeq\frac{\delta\chi}{\dot{\chi}},
\end{eqnarray}
corresponding to adiabatic perturbations, which is relevant when conditions (\ref{SMALL}) are satisfied.

Substituting eq. (\ref{ADIABATIC}) into (\ref{CURV}) we have
\begin{eqnarray}
\label{CURV4}
&&{\mathcal R}\simeq H\frac{\delta\phi}{\dot{\phi}}=H\frac{\delta\chi}{\dot{\chi}},
\end{eqnarray}
and, making substitution eq. (\ref{ADIABATIC}) into (\ref{ISOCURV}), we obtain
\begin{eqnarray}
\label{ISOCURV4}
&&{\mathcal S}\simeq 0.
\end{eqnarray}

In addition, we note that from (\ref{ISOCURV3}), (\ref{ADIABATIC}) and (\ref{ISOCURV4}) we get
\begin{eqnarray}
\label{ADIABATIC2}
&&\frac{\delta\phi}{\dot{\phi}}\simeq\frac{\delta\chi}{\dot{\chi}}\simeq\frac{\delta\theta}{\dot{\theta}},
\end{eqnarray}
what is a consequence of the linear relationship (\ref{RELDS}) as well.

Summimg up, we can state that in proposed multifield cosmological models the isocurvature perturbations are negligible, and observable curvature perturbations ${\mathcal R}$ are induced by adiabatic modes only.
Thus, in this situation, the analysis of cosmological perturbations (in a linear order) can be reduced to perturbations of a single scalar field, and expressions (\ref{PSGB})--(\ref{P7}) correspond to correct parameters of cosmological perturbations. Therefore they can be used to verify the proposed inflationary models on the basis of observational constraints (\ref{PLANCK1})--(\ref{PLANCK2}).

\section{The examples of phantonical models based of exact inflationary solutions}\label{EXACT-CC}

In this section we consider few examples of exact cosmological solutions for phantonical inflationary models based on proposed here approach. We will analyze asymptotic behavior of obtained exact inflationary solutions to consider the possibility of realizing the observed second accelerated expansion of the universe in the framework of the model closely to $\Lambda$CDM--model \cite{Copeland:2006wr}. It is well-known that $\Lambda$CDM--model is in a good agreement with observational data for current stage of the universe's evolution \cite{SupernovaCosmologyProject:1998vns,SupernovaSearchTeam:1998fmf}.

To describe the second universe acceleration, we request that at large times $t\gg t_{i}$, the exact inflationary solutions should be reduced to (\ref{A1})--(\ref{A2}) and (\ref{A4}), which correspond to the effective cosmological constant. The remaining material components that have a significant effect on the dynamics of the universe in $\Lambda$CDM--model are Cold Dark Matter and Baryonic Matter\footnote{The influence of the radiation on the cosmological dynamic is negligible compared to these components at present era \cite{Planck:2018vyg}.}, and the source of the second accelerated expansion of the universe is effective cosmological constant $\Lambda_{eff}$.
An approximate expression for the effective cosmological constant (\ref{A4}) corresponds to the small deviation in observational estimates of the Dark Energy state parameter $w_{DE}=-1.03\pm0.03$ from the value $w_{\Lambda}=-1$ \cite{Planck:2018vyg}.

It is worth to note here that the combination of observational estimates of the values of the Hubble parameter at the present time $H_{0}$, Planck CMB power-spectrum data, and BAO data constraint makes it possible to estimate the relative density of matter $\Omega_{m}=\rho_{m}/\rho_{c}$ and, accordingly, the relative density of Dark Energy $\Omega_{\Lambda}=\rho_{\Lambda}/\rho_{c}$. It gives the following estimate of the value of the cosmological constant in $\Lambda$CDM--model~\cite{Planck:2018vyg}
\begin{eqnarray}
\label{DARKEN}
&&\Lambda_{obs}=(2.846\pm0.076)\times10^{-122},
\end{eqnarray}
in natural system of units implying $M^{-2}_{p}=1$.
This condition, combined with observational restrictions on the values of the parameters of cosmological perturbations (\ref{PLANCK1})--(\ref{PLANCK2}), will allow us to determine the admissible values of the constant parameters of cosmological models.

As examples of the application of the proposed approach, we will use well-known types of cosmological dynamics, considering earlier for one-field models \cite{Chervon:2019sey,Chervon:2017kgn}, to construct the exact inflationary solutions for phantonical two-field models and analyse these solutions at the large times.

\subsection{The exact solutions for the power-law inflation}\label{PLHINF}

Now, we consider the power-law inflation with the following scale factor
\begin{equation}
\label{PLI}
a(t)=a_{0}t^{m},
\end{equation}
where $m>1$ is a constant.

Substituting the scale factor  from (\ref{PLI}) into (\ref{PH1})--(\ref{PS1}) we find
\begin{equation}
\label{PLI1}
\phi(t)=\frac{\alpha}{\sqrt{2\omega}}\ln\left(\frac{A}{m}\,t^{m+1}\right),
\end{equation}
\begin{equation}
\label{PLI2}
\chi(t)=\frac{\beta}{\sqrt{2\omega}}\ln\left(Bm\,t^{m-1}\right).
\end{equation}

Similarly, from (\ref{EX1}) and (\ref{EX3}) one can get
\begin{eqnarray}
\label{PLN1}
&&N(t)=m\ln t,\\
\label{PHAS}
&&H(t)=\frac{m}{t},
\end{eqnarray}
and after substitution (\ref{PLN1}) into (\ref{EX6}) we have
\begin{equation}
\label{PLN2}
F(N(t))=(3m-1)t^{-1}.
\end{equation}
Taking into account (\ref{PLN1}), we obtain
\begin{equation}
\label{PLN3}
F(N)=(3m-1)\exp\left(-\frac{N}{m}\right).
\end{equation}

Using the condition $N=0$ we obtain the following time of the beginning of inflation: $t=t_{i}=1$, and
from (\ref{PLN1}) we get $\dot{N}(t=t_{i})=m$.
Further, from (\ref{PLN3}) one has $F(N=0)=3m-1$, thus the model satisfy both conditions (\ref{CR1})--(\ref{CR2}).

Substituting (\ref{PLN3}) into (\ref{EX15}) and using (\ref{EX9}),  we obtain
\begin{equation}
\label{PLI7A}
V(\phi,\chi)=A(3m-1)\left(AB\right)^{\frac{1-m}{2m}}\times
\exp\left(-\sqrt{\frac{\omega}{2}}\frac{(m+1)}{m\alpha}\phi
+\sqrt{\frac{\omega}{2}}\frac{(m-1)}{m\beta}\chi\right).
\end{equation}

It is important to note that similar solutions for the power-law inflationary model with two canonical fields were considered earlier in \cite{Elizalde:2004mq}.

Under condition $m\gg1$ we have
\begin{equation}
\label{PLI7}
V(\phi,\chi)\simeq 3m\sqrt{\frac{A}{B}}\exp\left(-\sqrt{\frac{\omega}{2}}\frac{\phi}{\alpha}
+\sqrt{\frac{\omega}{2}}\frac{\chi}{\beta}\right).
\end{equation}

In addition, for $m\gg1$ from (\ref{PLI1})--(\ref{PLI2}) we obtain
\begin{eqnarray}
\label{ALI1}
&&\phi(t)\simeq\frac{\alpha}{\sqrt{2\omega}}\left[\ln t+\ln\left(\frac{A}{m}\right)\right],\\
\label{ALI2}
&&\chi(t)\simeq\frac{\beta}{\sqrt{2\omega}}\left[\ln t+\ln\left(Bm\right)\right],
\end{eqnarray}
and substituting (\ref{ALI1})--(\ref{ALI2}) into (\ref{PLI7}) we get
\begin{equation}
\label{PLI8}
\Lambda_{eff}=V(m\gg1)\simeq3m^{2}.
\end{equation}

Thus, the model of power-law inflation under condition $m\gg1$ is closely to the effective cosmological constant for any cosmic time $t$.

The slow-roll parameters (\ref{SR1}) corresponding to Hubble parameter $H(t)=m/t$ for this model are
\begin{eqnarray}
\label{PLI9}
&&\epsilon=\delta=\frac{1}{m}.
\end{eqnarray}

For $m\gg1$ the slow-roll conditions (\ref{SR1}) are satisfied.

For the scale factor (\ref{PLI}) we have corresponding Hubble parameter $H=m/t$, and from (\ref{P2})--(\ref{P4}) we obtain the exact solution for canonical scalar field
\begin{eqnarray}
\label{PLI10}
&&\sigma(t)=\sqrt{2m}\ln t,\\
\label{PLI11}
&&V(\sigma)=m(3m-1)\exp\left(-\sqrt{\frac{2}{m}}\sigma\right),
\end{eqnarray}
corresponding to the power-law inflation, which was considered earlier in \cite{Ivanov,Lucchin:1984yf}.
The parameters of cosmological perturbations for this model based on expressions (\ref{PSGB})--(\ref{SR1}) were considered earlier in \cite{Fomin:2019yls}.

Using expressions (\ref{P5}) and (\ref{P7}) for models' slow-roll parameters (\ref{PLI9}) we obtain
\begin{eqnarray}
\label{PLI10INF}
&&r=\frac{4s(n_{S}-3)}{(n_{S}-1)}.
\end{eqnarray}

\begin{figure}[ht]
\centering
\includegraphics[width=9 cm]{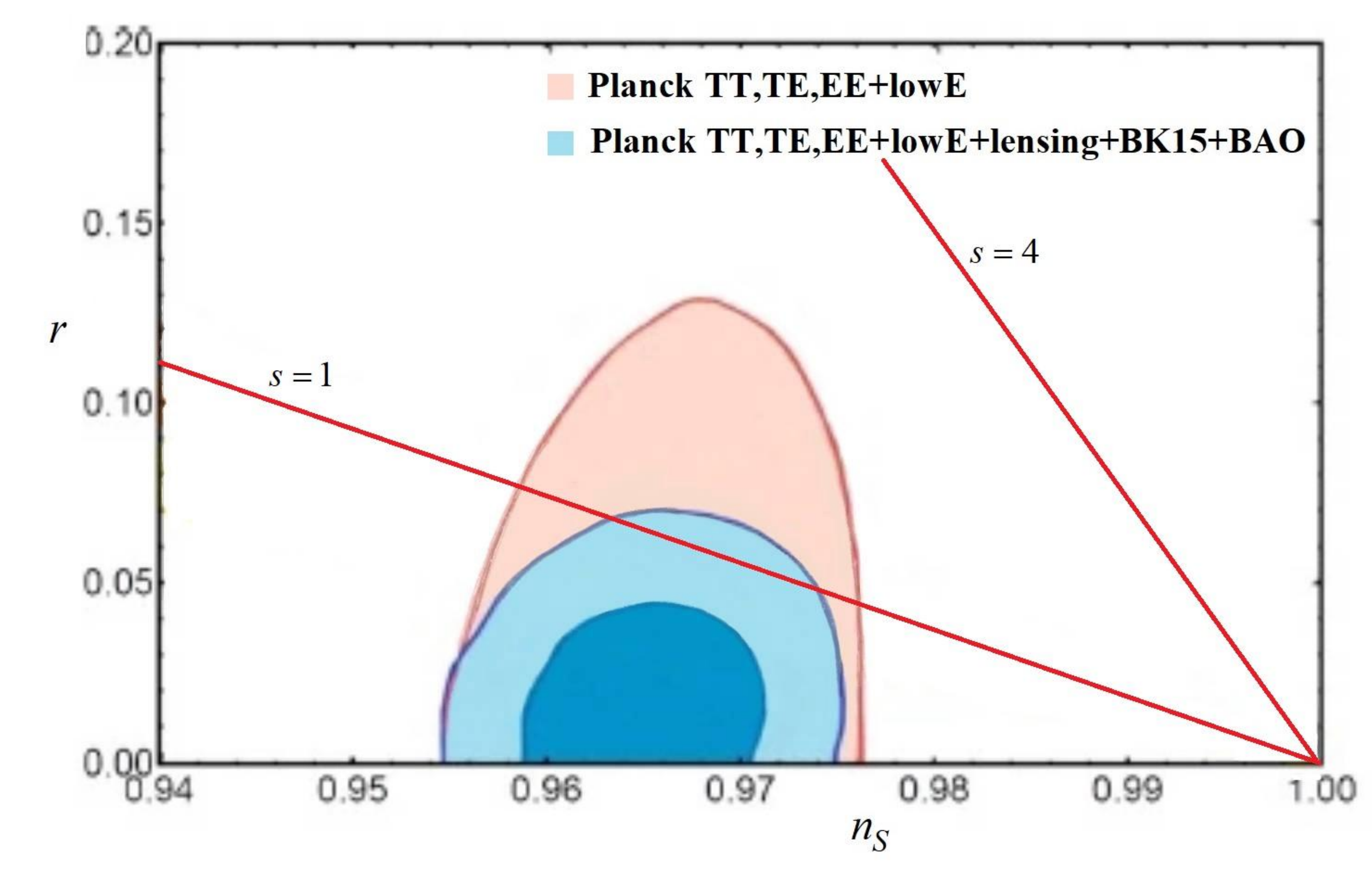}
\caption{The dependence $r=r(n_{S})$ for different values of the parameter $s=1,4$ with constraints on the tensor-to-scalar ratio $r_{0.002}$ due to the Planck TT,TE,EE+lowE (red) and the Planck TT,TE,EE+lowE+lensing+BAO+BICEP2/Keck Array (blue) observations~\cite{Planck:2018vyg,BICEP2:2018kqh}.}
\label{fig1PL}
\end{figure}

 Fig.\ref{fig1PL} represents graphic of the dependence $r$ on $n_s$ (\ref{PLI10INF}) for the inflationary model with Hubble parameter (\ref{PHAS}).
 As a criterion for verifying inflationary models according to observational constraints on the values of the parameters of cosmological perturbations, we take the values of the tensor-to-scalar ratio $r$ and the spectral index of scalar perturbations $n_{S}$ which fall into the outer and inner blue regions corresponding to 68\% and 95\% confidence level.

The power-law inflationary model (\ref{PLI})-(\ref{PLN3}) can be verified by the observational constraints (\ref{PLANCK1})--(\ref{PLANCK2}) for $s=1$
(with 68\% confidence level) and it doesn't correspond these constraints for $s=4$.
Thus, the verification of this inflationary model by the observational constraints on the parameters of cosmological perturbations depends on the value of parameter $s$.

Substituting (\ref{PLI9}) into (\ref{P5}) and taking into account constraints (\ref{PLANCK2}), one has the following restrictions on the constant parameter of the model $53\leq m\leq68$.

In addition, using the expressions (\ref{PLN1})--(\ref{PHAS}) and (\ref{PLI8})--(\ref{PLI9}) one can find
\begin{eqnarray}
\label{COSMHPL}
&&A_{S}\simeq\frac{m\Lambda_{eff}}{24\pi^{2}}\exp\left(-\frac{2N}{m}\right),
\end{eqnarray}
where $A_{S}=2.1\times10^{-9}$ and $N\simeq60$ on the crossing of the Hubble radius.


 Under condition $53\leq m\leq68$ from (\ref{COSMHPL}) we find for the power-law inflation the following constraints on the effective cosmological constant:
 $$
 4\times10^{-8}\leq\Lambda_{eff}\leq9\times10^{-8}.
$$

Thus, taking into account (\ref{DARKEN}) we obtain negative cosmological constant associated with vacuum energy $|\Lambda_{vac}|\simeq\Lambda_{eff}$.

\subsection{The exact solutions for generalized hyperbolic inflation}\label{hyp}

 We turn, now, to the analysis of cosmological dynamics based on the scale factor determined by hyperbolic sine and hyperbolic cosine in the framework of the inflationary models.
 In the general case, this type of cosmological dynamics differs from the previously considered solutions (\ref{COSH})--(\ref{SINH}) and is reduced to them in a particular case.

\subsubsection{The dynamics defined by hyperbolic sine}

Let us consider an inflationary model with the following scale factor
\begin{equation}
\label{HS}
a(t)=a_{0}\sinh^{k}(\lambda t),
\end{equation}
where $\lambda>0$ and $k>0$ are constants.
The corresponding e-folds number and its derivative are
\begin{eqnarray}
\label{HS1}
&&N(t)=k\ln \left[\sinh(\lambda t)\right],\\
\label{HS2}
&&\dot{N}=H(t)=k\lambda\,\frac{\cosh(\lambda t)}{\sinh(\lambda t)}.
\end{eqnarray}

Using the condition $N=0$ we have
\begin{equation}
\label{HS3}
t_{i}=\frac{1}{\lambda}\ln\left(1+\sqrt{2}\,\right),
\end{equation}
and, therefore, we obtain $\dot{N}(t=t_{i})=\sqrt{2}k\lambda\neq0$.

Substituting (\ref{HS1}) into (\ref{EX6}) one can get
\begin{equation}
\label{HS4}
F(N)=\frac{\lambda\left[3k e^{N/k}+(3k-1)e^{-N/k}\right]}
{\left(1+e^{2N/k}\right)^{1/2}}.
\end{equation}
If $N=0$, eq. (\ref{HS4}) leads to
\begin{equation}
\label{HS5}
F(N=0)=\frac{\lambda}{\sqrt{2}}(6k-1).
\end{equation}
Thus, the model (\ref{HS}) satisfy both conditions (\ref{CR1})--(\ref{CR2}).

Using equations (\ref{PH1})--(\ref{PS1}) or (\ref{EX7})--(\ref{EX8}) we can find
\begin{equation}
\label{HS6}
\phi(t)=\frac{\alpha}{\sqrt{2\omega}}\ln\left[\frac{A}{k\lambda}\frac{\sinh^{k+1}(\lambda t)}
{\cosh(\lambda t)}\right],
\end{equation}
\begin{equation}
\label{HS7}
\chi(t)=\frac{\beta}{\sqrt{2\omega}}\ln\left[Bk\lambda\sinh^{k-1}(\lambda t)\cosh(\lambda t)\right].
\end{equation}
The exact expression for the potential is
\begin{equation}
\label{HS8}
V(\phi,\chi)=\lambda\sqrt{\frac{A}{B}}\exp\left(-\sqrt{\frac{\omega}{2}}\frac{\phi}{\alpha}
+\sqrt{\frac{\omega}{2}}\frac{\chi}{\beta}\right)
\frac{\left[3k e^{\frac{N(\phi,\chi)}{k}}+(3k-1)e^{-\frac{N(\phi,\chi)}{k}}\right]}
{\left(1+e^{\frac{2N(\phi,\chi)}{k}}\right)^{1/2}},
\end{equation}
where
\begin{equation}
\label{HS8A}
N(\phi,\chi)\equiv\sqrt{\frac{\omega}{2}}\frac{\phi}{\alpha}
+\sqrt{\frac{\omega}{2}}\frac{\chi}{\beta}-\frac{1}{2}\ln(AB),
\end{equation}
due to eq. (\ref{EX9}).

Under condition $e^{N/k}\gg1$ (corresponding to $t\gg t_{i}$) from (\ref{HS4}) we have the following expression for the potential
\begin{equation}
\label{HS9}
V(\phi,\chi)\simeq3k\lambda\sqrt{\frac{A}{B}}\exp\left(-\sqrt{\frac{\omega}{2}}\frac{\phi}{\alpha}
+\sqrt{\frac{\omega}{2}}\frac{\chi}{\beta}\right).
\end{equation}
Besides, under condition $t\gg t_{i}$ we get
\begin{eqnarray}
\label{HSA}
&&\sinh(\lambda t)\simeq\cosh(\lambda t)\simeq\frac{1}{2}\exp(\lambda t),
\end{eqnarray}
and from (\ref{HS6})--(\ref{HS7}) one can obtain
\begin{eqnarray}
\label{AL1A}
&&\phi(t)\simeq\frac{\alpha}{\sqrt{2\omega}}\left[k\lambda t+\ln\left(\frac{A}{k\lambda}\right)\right],\\
\label{AL2A}
&&\chi(t)\simeq\frac{\beta}{\sqrt{2\omega}}\left[k\lambda t+\ln\left(k\lambda B\right)\right].
\end{eqnarray}
Substituting (\ref{AL1A})--(\ref{AL2A}) into (\ref{HS9}) we have
\begin{equation}
\label{HS10}
\Lambda_{eff}=V(t\gg t_{i})\simeq3k^{2}\lambda^{2},
\end{equation}
i.e. in this model the effective cosmological constant depends on two parameters.

Thus, the model (\ref{HS}) is closely to the $\Lambda$CDM--model of the second accelerated stage of the universe evolution at the times $t\gg t_{i}$.

The slow-roll parameters (\ref{SR1}), corresponding to Hubble parameter (\ref{HS2}) for this model, are
\begin{eqnarray}
\label{HSI9}
&&\epsilon=\frac{1}{k\cosh^{2}(\lambda t)},~~~~
\delta=\frac{1}{k},
\end{eqnarray}
It is clear, if $k\gg1$ the slow-roll conditions (\ref{SR1}) are satisfied.

Using Hubble parameter in the form (\ref{HS2}) and eqs. (\ref{P2})--(\ref{P4}) we obtain the exact solution
\begin{eqnarray}
\label{HSI10}
&&\sigma(t)=\pm\sqrt{2k}\ln\left[\tanh\left(\frac{\lambda}{2}t\right)\right],\\
\label{HSI11}
&&V(\sigma)=\frac{k\lambda^{2}}{2}\left[3k+1+(3k-1)\cosh\left(\sqrt{\frac{2}{k}}\sigma\right)\right],
\end{eqnarray}
for corresponding single filed inflationary model.

Let us note, this model was firstly considered in \cite{Barrow:1994nt}, and subsequently in \cite{Motohashi:2014ppa} in the context of the constant-roll inflationary models.

Taking into account the slow-roll parameters (\ref{HSI9}) and using  (\ref{P5}) and (\ref{P7}), we find the following dependence tensor-to-scalar ratio on spectral index of scalar perturbations
\begin{eqnarray}
\label{HSI12}
&&r=\frac{4s(kn_{S}-k-2)}{k(n_{S}-5)}.
\end{eqnarray}

If $s=4$ under conditions (\ref{PLANCK1})--(\ref{PLANCK2}) we have negative values of the constant $k<0$ which contradicts to condition $\epsilon>0$. If $s=1$ under conditions (\ref{PLANCK1})--(\ref{PLANCK2}) we get the following restriction on the constant $k$: $k>56$.

If $k\gg1$ from (\ref{HSI12}) under condition $\frac{2}{k}\ll|n_{S}-1|$ we obtain the following dependence
\begin{eqnarray}
\label{HSI13}
&&r\simeq\frac{4s(n_{S}-1)}{(n_{S}-5)},
\end{eqnarray}
corresponding to the minimal value of the tensor-to-scalar ratio.

\begin{figure}[ht]
\centering
\includegraphics[width=9 cm]{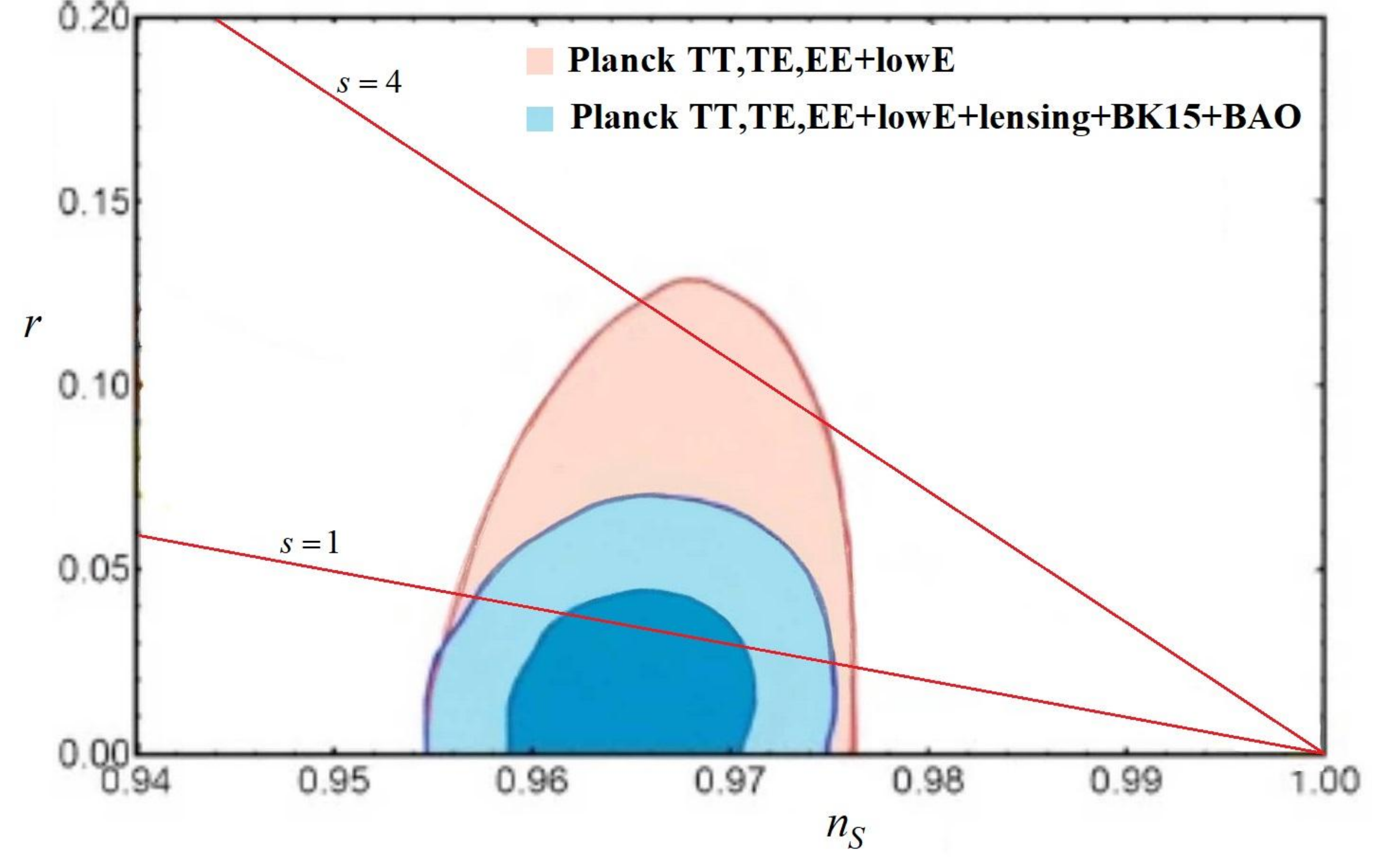}
\caption{The dependence $r=r(n_{S})$ for $k\gg1$ and different values of the parameter $s$ with constraints on the tensor-to-scalar ratio $r_{0.002}$  due to the Planck TT,TE,EE+lowE (red) and the Planck TT,TE,EE+lowE+lensing+BAO+BICEP2/Keck Array (blue) observations~\cite{Planck:2018vyg,BICEP2:2018kqh}.}
\label{fig1}
\end{figure}

In the Fig.\ref{fig1} the dependence (\ref{HSI13}) for the inflationary model with Hubble parameter (\ref{HS2}) is represented.
As one can see, the model doesn't correspond to the observational constraints (\ref{PLANCK1})--(\ref{PLANCK2}) for $s=4$, however, it can be verified by these constraints on the parameters of cosmological perturbations for $s=1$.

Now, we estimate the value of the cosmological constant for this inflationary model based on the value of the power spectrum of scalar perturbations (\ref{PLANCK1}).

Using eqs. (\ref{HS1})--(\ref{HS2}) and (\ref{HSI9}) we get
\begin{eqnarray}
\label{HSI14}
&&H^{2}=k^{2}\lambda^{2}\left(1+e^{-2N/k}\right),\\
\label{HSI15}
&&\epsilon=\frac{1}{k}\left(1+e^{2N/k}\right)^{-1}.
\end{eqnarray}
Substituting (\ref{HSI14}) and (\ref{HSI15}) into (\ref{PSGB}) we find
\begin{eqnarray}
\label{HSI16}
&&A_{S}=\frac{k^{3}\lambda^{2}}{4\pi^{2}}\left[1+\cosh\left(\frac{2N}{k}\right)\right]\simeq
\frac{k\Lambda_{eff}}{12\pi^{2}}\left[1+\cosh\left(\frac{2N}{k}\right)\right],
\end{eqnarray}
where $A_{S}=2.1\times10^{-9}$ and $N\simeq60$ on the crossing of the Hubble radius.

Using (\ref{HSI16}) under condition $k>56$ for this model one has the following constraint on the parameter $\lambda $: $0<\lambda<3\times10^{-7}$, and on the effective cosmological constant $\Lambda_{eff}$: $0<\Lambda_{eff}<8\times10^{-10}$.

Thus, we have the following cosmological constant associated with vacuum energy $\Lambda_{vac}=\Lambda_{obs}-\Lambda_{eff}$, which is positive $\Lambda_{vac}>0$ for $\Lambda_{eff}<\Lambda_{obs}$, negative $\Lambda_{vac}<0$ for $\Lambda_{eff}>\Lambda_{obs}$,  and it is equal to zero $\Lambda_{vac}=0$ for $\Lambda_{eff}=\Lambda_{obs}$, where the observed cosmological constant $\Lambda_{obs}$ is defined in (\ref{DARKEN}).

\subsubsection{The dynamics defined by hyperbolic cosine}

Considering the model with the scale factor
\begin{equation}
\label{HC1}
a(t)=a_{0}\cosh^{k}(\lambda t),
\end{equation}
one can find
\begin{equation}
\label{HC2}
N(t)=k\ln \left[\cosh(\lambda t)\right],
\end{equation}
and, substituting (\ref{HC2}) into (\ref{EX6}), we get
\begin{equation}
\label{HC4}
F(N)=\frac{\lambda\left[3k e^{N/k}+(1-3k)e^{-N/k}\right]}
{\left(e^{2N/k}-1\right)^{1/2}}.
\end{equation}

 If $N=0$, one has $F(N)\rightarrow\infty$ and $V\rightarrow\infty$, thus, the dynamics for the scale factor (\ref{HC1}) doesn't correspond to condition (\ref{CR2}) and we exclude this case from further consideration.

\subsection{\label{sec:ghi}The exact solutions for exponential power-law inflation}

Considering the case of the exponential power-law inflation we define the scale factor as follows
\begin{equation}
\label{PLE}
a(t)=a_{0}t^{m}\exp(\lambda t),
\end{equation}
where $\lambda>0$ and $m>0$ are some positive constants.

The corresponding e-folds number and its derivative are
\begin{eqnarray}
\label{PLE1}
&&N(t)=\ln\left[t^{m}\exp(\lambda t)\right],\\
\label{PLE2}
&&\dot{N}=H(t)=\lambda+\frac{m}{t}.
\end{eqnarray}
From the condition $N=0$ we have
\begin{equation}
\label{PLE3}
t_{i}=\frac{m}{\lambda}\,W\left(\frac{\lambda}{m}\right),
\end{equation}
where $W$ denotes the Lambert function \cite{Abramowitz}.

Thus, we get
\begin{equation}
\label{PLE4}
\dot{N}(t=t_{i})=\frac{\lambda\left[1+W\left(\frac{\lambda}{m}\right)\right]}
{W\left(\frac{\lambda}{m}\right)}.
\end{equation}

Substituting (\ref{PLE1}) into (\ref{EX6}) we obtain
\begin{equation}
\label{PLE5}
F(N)=\frac{3\lambda^{2}+6\lambda m \exp\left[-\frac{N}{m}+W\left(\frac{\lambda}{m}e^{\frac{N}{m}}\right)\right]+
m(3m-1)\exp\left[-2\frac{N}{m}+2W\left(\frac{\lambda}{m}e^{\frac{N}{m}}\right)\right]}
{\lambda+m \exp\left[-\frac{N}{m}+W\left(\frac{\lambda}{m}e^{\frac{N}{m}}\right)\right]}.
\end{equation}
If $N=0$, we get
\begin{equation}
\label{PLE6}
F(N=0)=\frac{\lambda\left[3W^{2}\left(\frac{\lambda}{m}\right)+6W\left(\frac{\lambda}{m}\right)+
\left(\frac{3m-1}{m}\right)\right]}
{W\left(\frac{\lambda}{m}\right)\left[1+W\left(\frac{\lambda}{m}\right)\right]}.
\end{equation}

Therefore, both conditions (\ref{CR1})--(\ref{CR2}) are satisfied for $\frac{\lambda}{m}>0$.

Using equations (\ref{PH1})--(\ref{PS1}) or (\ref{EX7})--(\ref{EX8}) we can derive
\begin{eqnarray}
\label{PLE7}
&&\phi(t)=\frac{\alpha}{\sqrt{2\omega}}\ln\left[\frac{Ae^{\lambda t}t^{m+1}}{\lambda t+m}\right],\\
\label{PLE8}
&&\chi(t)=\frac{\beta}{\sqrt{2\omega}}\ln\left[Be^{\lambda t}\left(mt^{m-1}+\lambda t^{m}\right)\right].
\end{eqnarray}

The exact expression for the potential is
\begin{equation}
\label{PLE9}
V(\phi,\chi)=\lambda\sqrt{\frac{A}{B}}\exp\left(-\sqrt{\frac{\omega}{2}}\frac{\phi}{\alpha}
+\sqrt{\frac{\omega}{2}}\frac{\chi}{\beta}\right)
\left[\frac{3\lambda^{2}+6\lambda m U(\phi,\chi)+m(3m-1)U^{2}(\phi,\chi)}{\lambda+m U(\phi,\chi)}\right],
\end{equation}
where
\begin{equation}
\label{PLE10}
U(\phi,\chi)\equiv\exp\left[-\frac{N(\phi,\chi)}{m}+W\left(\frac{\lambda}{m}e^{\frac{N(\phi,\chi)}{m}}\right)\right],
\end{equation}
and
\begin{equation}
\label{PLE11}
N(\phi,\chi)\equiv\sqrt{\frac{\omega}{2}}\frac{\phi}{\alpha}
+\sqrt{\frac{\omega}{2}}\frac{\chi}{\beta}-\frac{1}{2}\ln(AB),
\end{equation}
according to eq. (\ref{EX9}).

Under condition $N/m\gg1$ (corresponding to $t\gg t_{i}$) one has $U(\phi,\chi)\ll1$ and from (\ref{PLE9}) we obtain the following expression for the potential
\begin{equation}
\label{HI9}
V(\phi,\chi)\simeq3\lambda\sqrt{\frac{A}{B}}\exp\left(-\sqrt{\frac{\omega}{2}}\frac{\phi}{\alpha}
+\sqrt{\frac{\omega}{2}}\frac{\chi}{\beta}\right).
\end{equation}

If $t\gg t_{i}$ and $t\gg \frac{m}{\lambda}$ from (\ref{PLE7})--(\ref{PLE8}) we obtain
\begin{eqnarray}
\label{APLE1}
&&\phi(t)\simeq\frac{\alpha}{\sqrt{2\omega}}\left[\lambda t+m\ln t+\ln\left(\frac{A}{\lambda}\right)\right],\\
\label{APLE2}
&&\chi(t)\simeq\frac{\beta}{\sqrt{2\omega}}\left[\lambda t+m\ln t+\ln\left(\lambda B\right)\right].
\end{eqnarray}
Therefore, the potential (\ref{HI9}) for the scalar fields (\ref{APLE1})--(\ref{APLE2}) is reduced to
\begin{equation}
\label{HI10}
\Lambda_{eff}=V(t\gg t_{i})\simeq3\lambda^{2}.
\end{equation}
 This result corresponds to the model closely to the $\Lambda$CDM--model of the second accelerated stage of the universe's evolution as well.

The slow-roll parameters (\ref{SR1}) corresponding to Hubble parameter (\ref{PLE2}) for this model are
\begin{eqnarray}
\label{PLEI9}
&&\delta=\frac{1}{\lambda t+m},~~~~\epsilon=m\delta^{2},
\end{eqnarray}
For the case $(\lambda t+m)\gg1$ the slow-roll conditions (\ref{SR1}) are satisfied.

In addition, analysing eqs. (\ref{P2})--(\ref{P4}) with Hubble parameter (\ref{PLE2}) we obtain the exact solution for canonical scalar field
\begin{eqnarray}
\label{PLEI10}
&&\sigma(t)=\sqrt{2m}\ln t,\\
\label{PLEI11}
&&V(\sigma)=3\lambda^{2}+6\lambda m\exp\left(-\frac{1}{2}\sqrt{\frac{2}{m}}\sigma\right)+
3m(m-1)\exp\left(-\sqrt{\frac{2}{m}}\sigma\right),
\end{eqnarray}
 corresponding to single filed inflationary model, which was considered earlier in \cite{Chervon:2019sey,Fomin:2019yls,Chervon:2017kgn,Fomin:2020caa}.

It was also shown in papers \cite{Fomin:2019yls,Fomin:2020caa} that this model satisfies the observational constraints on the values of the parameters of cosmological perturbations (\ref{PLANCK1})--(\ref{PLANCK2}), regardless of the choice of parameter $s=1$ or $s=4$.

Substituting (\ref{PLEI9}) into  (\ref{P5}) and (\ref{P7}) and taking into account slow-roll conditions $\epsilon,\delta\ll1$ we find the dependence tensor-to-scalar ratio on spectral index of scalar perturbations as follows
\begin{eqnarray}
\label{PLEI12}
&&r\simeq sm(n_{S}-1)^{2}.
\end{eqnarray}

We can consider the case $s=4$ only, since if the inflationary model is verifiable for $s=4$, it is verifiable for $s=1$ as well.
Under conditions (\ref{PLANCK1})--(\ref{PLANCK2}) we have the restriction on the constant parameter $m$: $0<m<18$.

Using eqs. (\ref{PLE1})--(\ref{PLE2}) and definition of the slow-roll parameters we find
\begin{eqnarray}
\label{PLEI13}
&&H^{2}=(\lambda+mU)^{2},\\
\label{PLEI14}
&&\epsilon=m\left(m+\frac{\lambda}{U}\right)^{-2},\\
\label{PLEI15}
&&\delta=\left(m+\frac{\lambda}{U}\right)^{-1},
\end{eqnarray}
where $U$ is defined by eq. (\ref{PLE10}).

Thus, from eqs. (\ref{PSGB}) and (\ref{PLEI13})--(\ref{PLEI14}) we have
\begin{eqnarray}
\label{PLEI16}
&&A_{S}=\frac{(\lambda+mU)^{4}}{8\pi^{2}mU^{2}}=\frac{m\lambda^{2}}{8\pi^{2}}\frac{(1+W)^{4}}{W^{2}},
\end{eqnarray}
where
\begin{eqnarray}
\label{PLEI17}
&&W=W\left(\frac{\lambda}{m}e^{\frac{N}{m}}\right),
\end{eqnarray}
$A_{S}=2.1\times10^{-9}$ and $N\simeq60$ on the crossing of the Hubble radius.

The condition $\lambda>0$, combined with the slow-roll conditions $\epsilon,\delta<10^{-2}$ in (\ref{PLEI14})--(\ref{PLEI16}), gives  the following constraints $8\times10^{-10}<m<5$ and $2\times10^{-10}<\lambda<4\times10^{-5}$, i.e.
we have the effective cosmological constant $1\times10^{-19}<\Lambda_{eff}<6\times10^{-9}$, and corresponding negative cosmological constant associated with vacuum energy $|\Lambda_{vac}|\simeq\Lambda_{eff}$.

\begin{figure}[ht]
\centering
\includegraphics[width=9 cm]{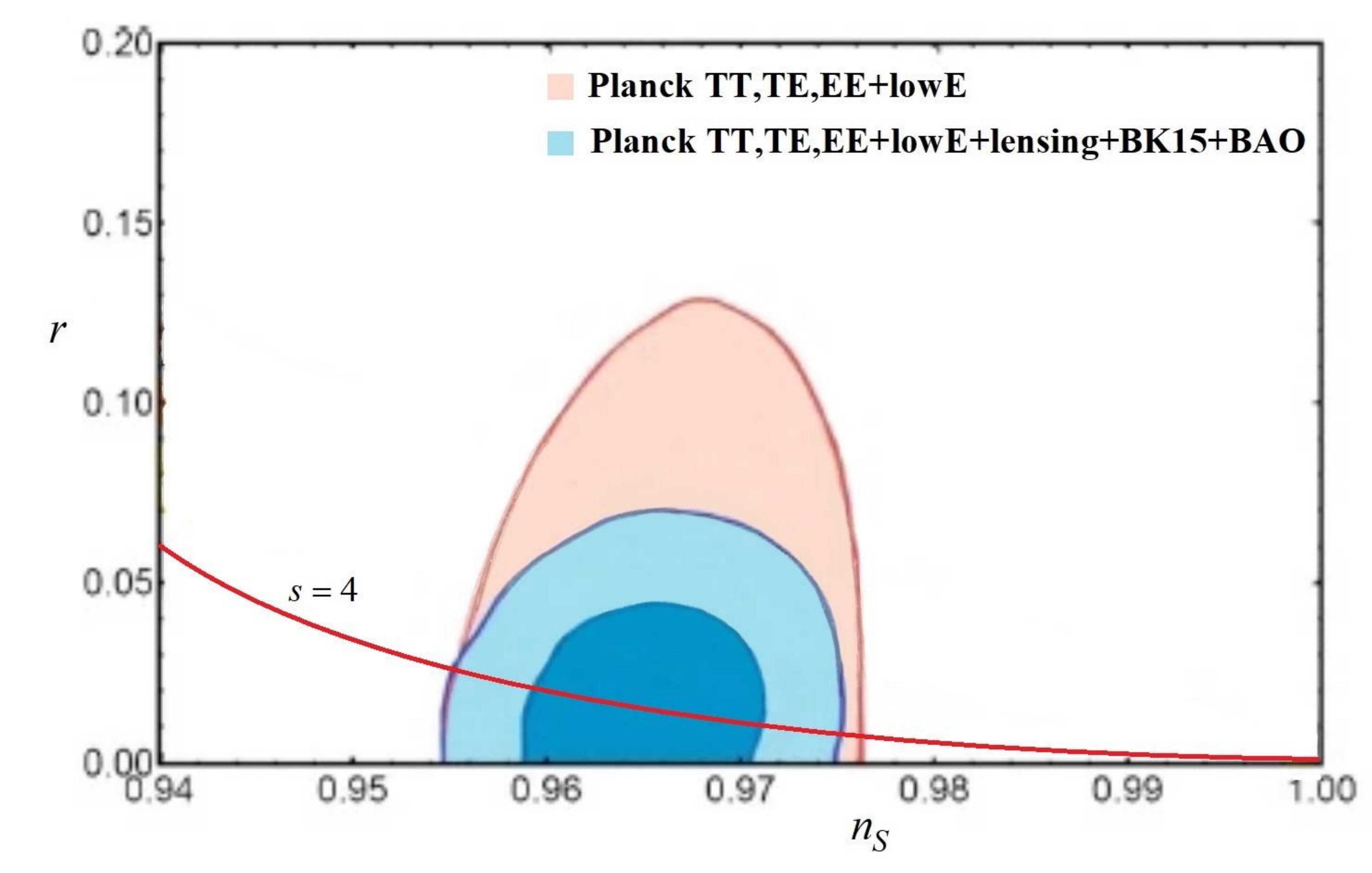}
\caption{The dependence $r=r(n_{S})$ for parameter $s=4$ and $m=4$ with constraints on the tensor-to-scalar ratio $r_{0.002}$ due to the Planck TT,TE,EE+lowE (red) and the Planck TT,TE,EE+lowE+lensing+BAO+BICEP2/Keck Array (blue) observations~\cite{Planck:2018vyg,BICEP2:2018kqh}.}
\label{fig1EPL}
\end{figure}

In Fig.\ref{fig1EPL} the dependence (\ref{PLEI12}) for the inflationary model with Hubble parameter (\ref{HS2}) is represented. As one can see, such a model corresponds to the observational constraints (\ref{PLANCK1})--(\ref{PLANCK2}) for $s=4$.


In addition, we note, that condition $m\ll1$ implies negligible tensor perturbations $r\simeq0$ in this model.

Despite the fact that theory of quantum gravity might allow different ground states with different values of the vacuum energy, as characteristic scale we consider $|\Lambda_{vac}|\simeq10^{-68}$ corresponding to the electro-weak phase transitions in early universe~\cite{Padmanabhan:2002ji,Martin:2012bt} in the natural system of units, which gives the following restriction
\begin{equation}
\label{CCPE1}
10^{-68}<|\Lambda_{vac}|\ll1,
\end{equation}
where $|\Lambda_{vac}|=1$ corresponds to the Planck energy scale.

As one can see, the estimates for the cosmological constant $\Lambda_{vac}$ associated with the quantum zero point fluctuations are within these constraints for proposed cosmological models.

\section{Conclusion}\label{conclusion}

We have considered two-filed cosmological models with canonical and phantom scalar fields as a chiral selfgravitating model \cite{Chervon:2020kdv} with weak kinetic and cross interactions between fields. I.e. the metric of the target space is pseudo-Euclidean and diagonal one.
It should be noted that most of the methods for generating exact cosmological solutions for the case of single scalar field models \cite{Chervon:2019sey} cannot be successfully applied to the considered two-field models.

First of all, we studied the model with a constant potential.
In this case fields' equations dictate linear dependence between fields.
This means that two-field phantonical model can be reduced to one-field model. We found the solutions for the case of a constant potential $V=const$, which are reduced to solutions for the case of one-field models.

Further we proposed new method of exact solutions constructions for two-filed phantonical models
with non-constant potential. This method we represented also in terms of e-fold number $N$ for the sake of easier cosmological parameters calculation.

It should be noted here that
methods for constructing exact solutions for multifield models \cite{Nojiri2006,Elizalde:2008yf,Elizalde:2004mq,Nojiri2015,Pali2014,Paliathanasis:2018vru,Anguelova:2018vyr,Chervon:2019nwq} considering a system with essential kinetic interaction (or non-trivial kinetic terms) could not be studied by our approach.
The superpotential method for two non-interacting scalar fields \cite{Vernov:2006dm} allow one to obtain the exact solutions for given scalar fields dependence on time. In our approach we fix scalar fields dependence on scale factor $a$ and its derivative $\dot{a}$.

Within the framework of the proposed approach, we postulate the direct connection between the type of cosmological dynamics (defined by a scale factor) and the evolution of scalar fields in two forms: (\ref{PH1})--(\ref{PS1}) -- in terms of a scale factor and (\ref{EX7})--(\ref{EX8}) -- in terms of e-fold number $N$.
If we suggest that dependencies $N=N(t)$ imply the inverse dependencies $t=t(N)$ in an explicit form, one can obtain exact analytical expressions for the potential $V(\phi,\chi)$ (\ref{V9}) and (\ref{EX15}).
Thus, one can define the exact solution for scalar fields and the potential for any, physically reasonable, scale factor $a=a(t)$ or corresponding e-folds number $N=N(t)$.

To be in physically reasonable situation,  we formulated restrictions (\ref{CR}) and (\ref{CR2}) on the possible types of cosmological dynamics from the conditions on scalar fields and their potential at the beginning of inflation.

 We note a very important property of the obtained solutions in quasi de Sitter approximation. Namely, we find that the potential (\ref{A3}) tends to the effective cosmological constant $\Lambda_{eff}$ at times significantly exceeding the time of the inflationary stage for various types of inflationary solutions. In such a way, this property allows us, using a model close to $\Lambda$CDM--model, to describe the observed later accelerated expansion of the universe in the present epoch. The effective cosmological constant in this case is inspired by the linear evolution of the doublet of the canonical and phantom fields, besides the value of the effective cosmological constant depends on the constant parameter of the model only.
Thus, based on the exact inflationary solutions for phantonical two-field models, one can select a class of models that leads to a later time accelerated expansion of the universe by virtue of the effective cosmological constant.

To study cosmological perturbation in the inflationary model under consideration we reduce two-field phantonical model to a single-field one via the relation (\ref{P1}).
It was also proven that isocurvature perturbations, in this case, do not have a significant effect on the parameters of cosmological perturbations. Moreover, these parameters can be calculated using corresponding single-field inflationary model.

The restrictions on the values of the constant parameters for proposed phantonical models were determined due to the present cosmological data. An estimate was also made of the cosmological constant associated with the quantum zero point fluctuations $\Lambda_{vac}$ from the value of the cosmological constant measured at the present time $\Lambda_{obs}$ and effective cosmological constant $\Lambda_{eff}$ for proposed models.

The observational constraints on the values of the parameters of cosmological perturbations limit the number of the inflationary models, since to assess their unconditional correctness, it is necessary to use a strict criterion $r=16\epsilon$ corresponding to $s=4$ in (\ref{PSGB}) and (\ref{P7}). From this point of view, only the exponential power-law inflationary model is verifiable among those considered in the paper.

However, one can consider generalized cosmological models, for example, taking into account the small corrections induced by the non-minimal coupling of the scalar field(s) with the Gauss-Bonnet term, so that arbitrary inflationary models can be verified by observational constraints on the parameters of cosmological perturbations for arbitrary value of the parameter $s$. In this case, observational constraint on the value of tensor-to-scalar ratio $r$ gives an estimate of the non-minimal coupling constant of the scalar field(s) and the Gauss-Bonnet term. This approach for a single-field cosmological models was demonstrated in \cite{Fomin:2020hfh}, and it can be further generalized for the proposed multi-field cosmological models.

It is important to note that one can obtain the exact solutions in phantonical two-field  models for the other types of inflationary dynamics and analyze the possibilities to construct phenomenological correct cosmological models in the framework of proposed approach.

\section{Acknowledgements}

Authors are thankful to reviewer of our work for comments and suggestion to include into consideration verification of constructed cosmological models. We think that corresponding extension of our work improved the article.

I.V.F. and S.V.C. were supported by the Russian Foundation for Basic Research grant No. 20-02-00280 a. and by the Russian Science Foundation grant No. 22-22-00248.  S.V.C. is grateful for the support by the Kazan Federal University Strategic Academic Leadership Program.

\end{document}